\documentclass[superscriptaddress,twocolumn,10pt,nofootinbib,floatfix]{revtex4-1}

\usepackage{amssymb}
\usepackage{amsmath}
\usepackage{appendix}
\usepackage{times}
\usepackage{graphicx}
\usepackage{subeqnarray}
\usepackage{bbold}
\usepackage{enumerate}
\usepackage{color}
\usepackage{bm}
\usepackage[colorlinks=true, letterpaper=true, pdfstartview=FitV, linkcolor=blue, citecolor=blue, urlcolor=blue]{hyperref}

\usepackage[normalem]{ulem}

\setcitestyle{super,open={},close={}}

\makeatletter
\renewcommand\@biblabel[1]{#1.}
\makeatother

\begin{document}
	
\title{Sliding Ferroelectric Metal with Ferrimagnetism}

\author{Zhenzhou Guo}
\affiliation{Institute for Superconducting and Electronic Materials, Faculty of Engineering and Information Sciences, University of Wollongong, Wollongong 2500, Australia.}

\author{Shifeng Qian}
\affiliation{Anhui Province Key Laboratory for Control and Applications of
Optoelectronic Information Materials, Department of Physics, Anhui Normal University, Wuhu, Anhui 241000, China.}

\author{Xiaodong Zhou}\thanks{Corresponding authors}\email{zhouxiaodong@tiangong.edu.cn}
\affiliation{School of Physical Science and Technology, Tiangong University, Tianjin 300387, China.}

\author{Wenhong Wang}
\affiliation{Institute of Quantum Materials and Devices, School of Electronic and Information Engineering, Tiangong University, Tianjin 300387, China.}

\author{Zhenxiang Cheng}\thanks{Corresponding authors}\email{cheng@uow.edu.au}
\affiliation{Institute for Superconducting and Electronic Materials, Faculty of Engineering and Information Sciences, University of Wollongong, Wollongong 2500, Australia.}

\author{Xiaotian Wang}\thanks{Corresponding authors}\email{xiaotianw@uow.edu.au}
\affiliation{Institute for Superconducting and Electronic Materials, Faculty of Engineering and Information Sciences, University of Wollongong, Wollongong 2500, Australia.}

	\begin{abstract}
		Two-dimensional (2D) sliding ferroelectric (FE) metals with ferrimagnetism represent a previously unexplored class of spintronic materials, featuring out-of-plane FE polarization, metallic conductivity, and a finite net magnetization, which together enable electrically tunable spintronic functionalities via FE switching. Here, based on antiferromagnetic (AFM) metallic bilayers, we propose a general strategy for constructing 2D sliding FE ferrimagnetic (FiM) metals that can achieve triply-coupled switching, in which the FE polarization, spin splitting, and net magnetization are reversed simultaneously through FE switching. As a prototypical realization, we design a bilayer sliding FE metal with FiM order, derived from monolayer Fe$_5$GeTe$_2$---a van der Waals metal with intrinsic ferromagnetic order close to room temperature. The system exhibits a FE transition from a nonpolar (NP) AFM phase to a FE FiM phase via interlayer sliding. The in-plane mirror symmetry breaking in FE metallic states lifts the nonrelativistic spin degeneracy that exists in the NP phase, leading to a sizable net magnetic moment. Furthermore, the interplay between metallicity, ferroelectricity, and ferrimagnetism gives rise to pronounced sign-reversible transport responses near the Fermi level, all of which can be electrically controlled by FE switching. Our results establish sliding FE metals with FiM as a promising platform for electrically reconfigurable, high-speed, and low-dissipation spintronic devices.
	\end{abstract}
	
	\maketitle

	
	Ferroelectric (FE) materials have long been studied for their rich physics and potential applications~\cite{buneTwodimensionalFerroelectricFilms1998,martinThinfilmFerroelectricMaterials2016,wu100YearsFerroelectricity2021,mengSlidingInducedMultiple2022,zhangFerroelectricOrderVan2023,SH-Li2024,huFerrielectricityControlledWidelytunable2024a,guoSpinPolarizedAntiferromagnetsSpintronics2025}. Among them, two-dimensional (2D) FE metals have attracted considerable attention due to their switchable out-of-plane FE polarization combined with metallic conductivity~\cite{sakaiCriticalEnhancementThermopower2016,huangPredictionIntrinsicFerromagnetic2018, feiFerroelectricSwitchingTwodimensional2018b,maLargeFamilyTwodimensional2021,yangTwoDimensionalTopologicalFerroelectric2024a}. This rare combination challenges the conventional understanding that ferroelectricity and metallicity are mutually exclusive, and opens up new opportunities for designing multifunctional devices in nanoelectronics and spintronics. However, most synthesized 2D materials intrinsically preserve inversion symmetry ($\mathcal{P}$) and do not support the FE property. Fortunately, sliding ferroelectricity, proposed in 2017~\cite{L-Li2017}, provides a viable mechanism to induce ferroelectricity in otherwise nonpolar (NP) multilayer systems via interlayer sliding, offering a pathway to 2D FE metals~\cite{yangOriginTwoDimensionalVertical2018a,houMetallicSlidingFerroelectricity2025}. To date, the intrinsically non-magnetic multilayer 1T'-WTe$_2$ remains the only experimentally confirmed system that simultaneously exhibits sliding ferroelectricity and metallicity~\cite{feiFerroelectricSwitchingTwodimensional2018b}, and integrating magnetism into such systems remains a major theoretical and experimental challenge.

	 Recent advances in the field of sliding ferroelectricity with collinear compensated magnetic order (including fully compensated ferrimagnets~\cite{liuMagnetoelectricCouplingMultiferroic2020a,KH-Liu2023,W-Xu2024,C-Wu2025} and altermagnets~\cite{YQ-Zhu2025,sunProposingAltermagneticFerroelectricTypeIII}) have still mainly focused on 2D insulating/semiconducting systems, which allow for the reversal of both FE polarization and spin splitting via FE switching. However, these systems maintain an open gap and zero net magnetization throughout the FE transition and lack metallic states near the Fermi level, limiting their potential for charge transport applications. In contrast, when collinear compensated magnetic order coexists with sliding ferroelectricity in metallic systems, the interplay among broken $\mathcal{P}$ (due to ferroelectricity) and time-reversal symmetry ($\mathcal{T}$) (due to magnetism), and Fermi-level metallic states can give rise to a sizable net magnetic moment and induce pronounced anomalous transport phenomena, including anomalous Hall effect (AHE), anomalous Nernst effect (ANE), and magneto-optical effects (MOEs). We refer to such systems as 2D sliding FE ferrimagnetic (FiM) metals. In these systems, electric-field-driven interlayer sliding enables triply-coupled switching---simultaneous control of the FE polarization, net magnetic moment, and spin-polarized band structures---offering a promising route toward nonvolatile tuning of anomalous transport properties. 2D sliding FE FiM metals that combine sliding ferroelectricity, ferrimagnetism, and giant, electrically switchable anomalous transport---preferably at room temperature---would mark a major milestone in materials design. However, neither such a design strategy nor corresponding materials have been reported to date, making their realization an important open challenge.

	\begin{figure}[htbp]
	\includegraphics[width=1\columnwidth]{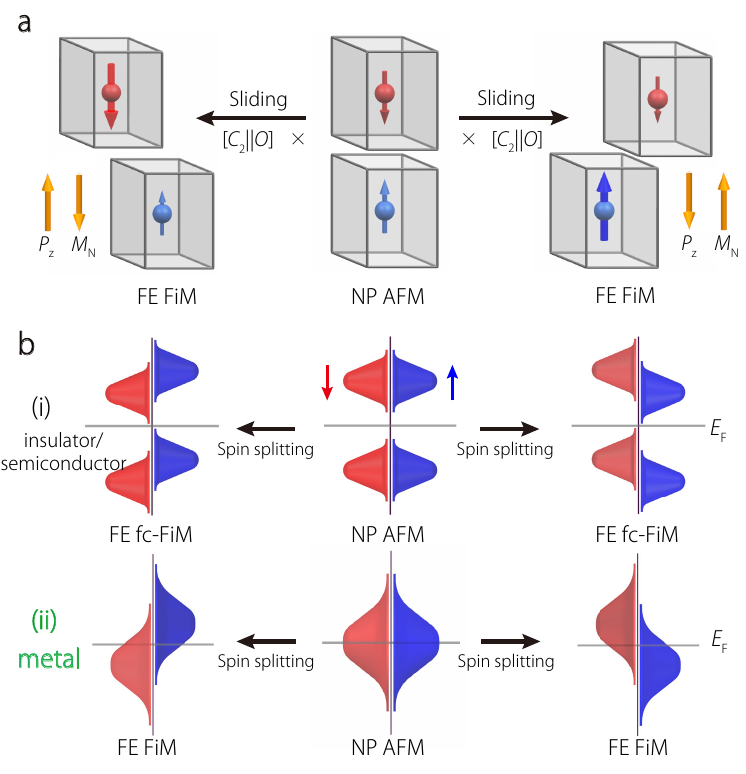}
	\caption{\textbf{Sliding FE magnetic insulators/semiconductors versus metals.} \textbf{a} Schematic illustration of the sliding FE transition from a NP AFM phase to a FE FiM phase. The red and blue arrows represent the atomic magnetic moments, respectively. The interlayer sliding breaks the spin group symmetry [$\mathcal{C}_{2}{\parallel}\mathcal{O}$] and induces a net magnetic moment and out-of-plane FE polarization, leading to a FE FiM phase. \textbf{b} Panels (i) and (ii) correspond to the insulating/semiconducting and metallic states, respectively. The red and blue shaded areas represent the spin-resolved DOS for the spin-down and spin-up channels, respectively. The net magnetization is given by the difference between $iDOS^\uparrow(E_F)$ and $iDOS^\downarrow(E_F)$.}
	\label{fig:fig1}
	\end{figure}

	In this paper, focusing on AFM bilayer metals, which possess nonrelativistic spin degeneracy under a $[\mathcal{C}_2||\mathcal{M}_z]$ symmetry operation, we propose a general strategy to construct the 2D sliding FE FiM metal, where the interplay of metallicity and lifted spin degeneracy leads to FiM order with a sizable net magnetic moment that can be efficiently modulated by FE switching. Taking bilayer Fe\textsubscript{5}GeTe\textsubscript{2} as a model system~\cite{mayFerromagnetismRoomTemperature2019,dengLayerNumberDependentMagnetismAnomalous2022,gopiThicknessTunableZoologyMagnetic2024}, we demonstrate that interlayer sliding induces a transition from a NP AFM state without net magnetization to a FE FiM phase with a significant net magnetization of up to 0.19~$\mu_B$ per unit cell, which exhibits a pronounced linear magnetoelectric coupling coefficient. Importantly, apart from out-of-plane FE polarization, the interlayer-sliding-driven FE switching can simultaneously switch the sign of the net magnetization and spin-splitting without altering the N{\'e}el vector. Such a triply-coupled switching can further enable reversible control of considerable anomalous transport properties by an electric field, opening new pathways for electrically tunable spintronic devices.

	\vspace{0.3cm}
	\noindent{\Large \bf Results}

	\vspace{0.05cm}	
	\noindent{\bf Triply-coupled switching in sliding FE FiM metals}

	\noindent
	Symmetry plays an essential role in understanding the properties of matter. The out-of-plane FE polarization in sliding FE FiM metals primarily arises from the potential difference between adjacent layers caused by interlayer sliding, which breaks the symmetry operation that connects the two opposite spins in the NP phase, such as $[\mathcal{C}_2||\mathcal{M}_z]$. This broken symmetry enables the simultaneous control of the FE polarization, net magnetic moment, and spin-polarized band structures, leading to a triply-coupled switching behavior.

	Here, we consider a FE transition from a NP A-type AFM  (intralayer ferromagnetic (FM) and interlayer AFM) phase to a FE FiM phase in a bilayer system, as schematically illustrated in Fig.~\ref{fig:fig1}a. We begin by constructing a bilayer NP phase from a monolayer ferromagnet. In the NP AFM phase, the two spin sublattices located in the top and bottom layers are connected by a combined spin symmetry operation, [$\mathcal{C}_2{\parallel}\mathcal{O}$], where $\mathcal{O}$ denotes a real-space symmetry operation relating opposite spin sublattices in different layers, such as $\mathcal{M}_z$. This operation guarantees both nonrelativistic spin degeneracy and a vanishing net magnetization. However, the combined operation [$\mathcal{C}_2{\parallel}\mathcal{O}$] only exists under specific stacking configurations. Interlayer sliding breaks this symmetry, leading to a transition into a FiM phase with spin-split band structures. Therefore, the key to identifying a suitable real material lies in the requirement that the stacking configuration, which is connected by the symmetry operation $\mathcal{O}$ relating the opposite spin sublattices of the two layers, must be breakable by interlayer sliding. Conversely, cases where $\mathcal{O} = \mathcal{P}$ must be excluded, as it cannot be broken by interlayer sliding. These conditions provide the necessary screening criteria for the material search.

	We then analyze the evolution of magnetization. In general, the net magnetization of a magnet is given by
\begin{equation}\label{eq:moment}
M = \mu_B \int_{-\infty}^{E_F} \left[ D^\uparrow(E) - D^\downarrow(E) \right] dE,
\end{equation}
where $\mu_B$ is the Bohr magneton, $E_F$ is the Fermi level, and $D^\uparrow(E)$ and $D^\downarrow(E)$ denote the spin-resolved density of states (DOS) for spin-up and spin-down electrons, respectively. Depending on electronic band structures of the system, two distinct scenarios can arise upon the FE transition (see Fig.~\ref{fig:fig1}b):

(i) For insulators/semiconductors, assuming the band gap remains open across the FE transition, it follows that the integrated DOS (iDOS) below the Fermi level consistently satisfies $iDOS^\uparrow(E_F) = iDOS^\downarrow(E_F)$. As such, even though the symmetry connecting the sublattices is broken during interlayer sliding, the integer filling condition ensures full magnetic compensation and a vanishing net magnetization, yielding a FE fully compensated ferrimagnetic (FE fc-FiM) insulating/semiconducting state~\cite{liuTwoDimensionalFullyCompensated2025}.

(ii) For metals, although the NP phase exhibits degenerate $iDOS$ due to the preserved nonrelativistic spin degeneracy, the sliding-induced FE transition lifts this degeneracy. As a result, the unequal values of $iDOS$ below the Fermi level ($iDOS^\uparrow(E_F) \neq iDOS^\downarrow(E_F)$) lead to a sizable net magnetic moment, referred to as FE FiM metal.

Based on this understanding, we propose a general strategy for realizing 2D sliding FE FiM metals. Starting from a bilayer (or multilayer) AFM metal constructed from certain monolayer metallic magnetic materials, interlayer sliding ferroelectricity is introduced to break the symmetry connecting the two spin sublattices. This symmetry breaking induces spin splitting at the Fermi level, leading to unequal $iDOS$ below the Fermi level ($iDOS^\uparrow(E_F) \neq iDOS^\downarrow(E_F)$), and consequently a finite net magnetization, forming a 2D sliding FE FiM metal.

Remarkably, in such systems, it is possible to achieve an interesting triply-coupled switching, i.e., simultaneous switching of the net magnetization $M_N$, the out-of-plane polarization $P_z$, and the spin splitting $\Delta E_{\mathbf{k}}^S$ (denoted by $S$)---without reversing the N{\'e}el vector---purely through interlayer sliding.
Considering an initial state $\psi_{\bm{k}}^{i}{(M_N, P_z, S)}$, one can always find a final state $\psi_{\bm{k}}^{f}{(-M_N, -P_z, -S)}$ that is related to the initial state by a combined operation $[\mathcal{C}_2||\mathcal{M}_z]$, i.e.,
\begin{equation}\label{eq:PT}
    [\mathcal{C}_2||\mathcal{M}_z] \, \psi_{\bm{k}}^{i}{(M_N, P_z, S)} = \psi_{\bm{k}}^{f}{(-M_N, -P_z, -S)}.
\end{equation}
Such a coupling between FE polarization and spin-polarized bands induced by ferrimagnetism, when further entangled with metallicity, can give rise to large anomalous transport properties that are controllable by an electric field.

\vspace{0.3cm}	
\noindent{\bf Material realization with triply-coupled switching}

\noindent
 To explore potential triply-coupled switching in real materials, we focus on a typically synthesized 2D metallic system, Fe$_5$GeTe$_2$, to construct a sliding FE structure. Fe$_5$GeTe$_2$ is a disordered vdW metal with intrinsic FM order near room temperature~\cite{mayFerromagnetismRoomTemperature2019,dengLayerNumberDependentMagnetismAnomalous2022,gopiThicknessTunableZoologyMagnetic2024}. Starting from the FM monolayer with the so-called UUU configuration, where the Fe5 atom is positioned above a Ge atom (see Supplementary Fig.~1 and Refs. \citenum{XX-Yang2021,gaoManipulationTopologicalSpin2022,schmittSkyrmionicSpinStructures2022,xieStrainInducedInterlayerMagnetic2023a,liPredictionStableNanoscale2024}), we considered eight distinct stackings: AA$^{\prime}$, AB, AC, BA, AA$^{\prime}$-1, AB-1, AC-1, and BA-1 (see Supplementary Fig.~2). As summarized in Supplementary Section A,  these eight stackings naturally fall into two sliding families, within each of which the four stackings can be transformed by in-plane sliding. The first family consists of AA$^\prime$, AB, AC, and BA, all of which favor A-type AFM ground state. In contrast, the second family includes AA$^\prime$-1, AB-1, AC-1, and BA-1, where AA$^\prime$-1 remains A-type AFM ground state, while its sliding counterparts (AB-1, AC-1, and BA-1) adopt FM ground state. Therefore, we primarily focus on the AA$^{\prime}$, AB, AC, and BA stackings, which exhibit
lower total energies and A-type AFM ground state, and
are most relevant to the present study.

	\begin{figure}
		\includegraphics[width=1.0\columnwidth]{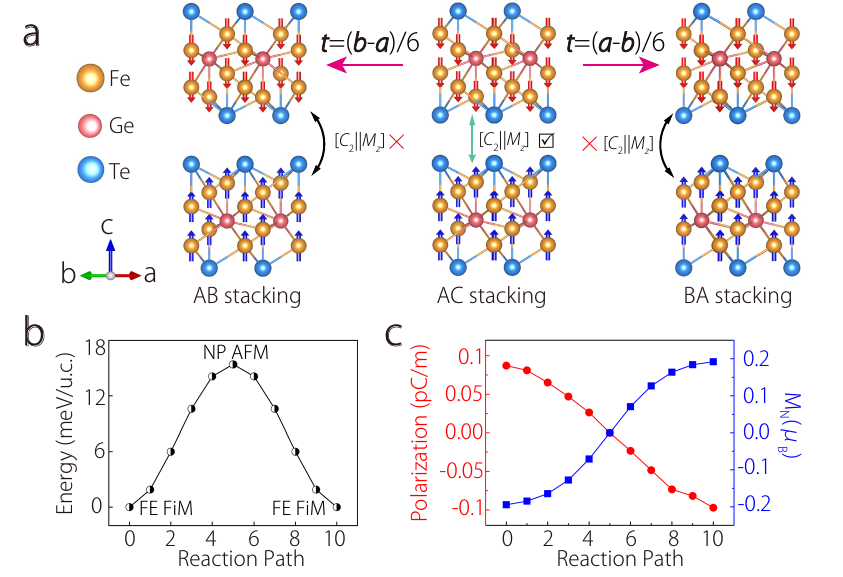}
		\caption{\textbf{Sliding configurations and FE switching in Fe$_5$GeTe$_2$ bilayer.} \textbf{a} Side views of Fe$_5$GeTe$_2$ bilayer with AB, AC, and BA stackings. The cyclic switching between three stackings can be achieved through the relative displacement vector $\bm{t}$ between layers. \textbf{b} Transition energy barriers in the FE switching path. \textbf{c} $P_z$ and $M_N$ as a function of the relative displacement vector $\bm{t}$.}
		\label{fig:fig2}
	\end{figure}

This bilayer structure of AC stacking belongs to space group $Aem2$ (No. 39), where the presence of the [$\mathcal{C}_2{\parallel}\mathcal{M}_z$] symmetry enforces a NP phase. By sliding the top layer relative to the bottom one along the in-plane vectors $\bm{t} = (\bm{b} - \bm{a})/6$ and $\bm{t} = (\bm{a} - \bm{b})/6$ (see Fig.~\ref{fig:fig2}a), the $M_z$ mirror symmetry is broken, resulting in two FE phases with degenerate energy but opposite $P_z$, corresponding to AB and BA stackings (No. 156, $P3m1$). The optimized lattice constants and atomic positions of AA$^{\prime}$, AB, BA, and AC stackings are listed in Supplementary Tables 1-4. The energy barrier of interlayer sliding between AB and BA stackings is estimated to be 15.4 meV per unit cell through the climbing image nudge-elastic-band (CL-NEB) method (see Fig.~\ref{fig:fig2}b). During the sliding process, although the relative atomic displacements within each layer are negligible~\cite{SH-Li2024}, the asymmetric atomic arrangement between the top and bottom layers induces an interlayer charge transfer, giving rise to a $P_z$~\cite{yangOriginTwoDimensionalVertical2018a,L-Li2017,xuBilayerJanusCr2I3Cl32025a}, which can be well defined by the classical formula in electrodynamics~\cite{yangTwoDimensionalTopologicalFerroelectric2024a}

	\begin{equation}\label{eq:polarization}
		P_z = \frac{1}{S} \int z\,\left( \rho_{\text{ions}} + \rho_{\text{valence}} \right) \, d^3r,
	\end{equation}
	where $S$ represents the area of the unit cell, and $\rho_{\text{ions}}$ and $\rho_{\text{valence}}$ denote the charge densities of ions and valence electrons, respectively. The $P_z$ of AB stacking is calculated to be 0.087 pC/m (see Fig.~\ref{fig:fig2}c). Considering the effective bilayer thickness (19.54 {\AA}) of Fe$_5$GeTe$_2$, this corresponds to a converted polarization of 44.5 $\mu\mathrm{C/m^2}$, which is comparable to those of other reported two-dimensional ferroelectrics, such as VCl$_2$ (23.6 $\mu\mathrm{C/m^2}$), VBr$_2$  (22.3 $\mu\mathrm{C/m^2}$), and VI$_2$ (12.2 $\mu\mathrm{C/m^2}$)~\cite{jiangFerroelectricityDrivenMagnetismMetal2025}, VAl$_2$S$_4$  (0.08 pC/m)~\cite{yuElectricalControlNoncollinear2024}, as well as 1T$^{\prime}$-ReS$_2$  (0.07 pC/m)~\cite{wanRoomTemperatureFerroelectricity12022}. Notably, ferroelectricity in 1T$^{\prime}$-ReS$_2$ has been experimentally confirmed using the piezoresponse force microscopy technique~\cite{wanRoomTemperatureFerroelectricity12022}, supporting the experimental feasibility of the predicted FE behavior in Fe$_5$GeTe$_2$. To further elucidate the charge transfer characteristics of the AB and BA stackings, Bader charges were calculated~\cite{xuBilayerJanusCr2I3Cl32025a}, as shown in Supplementary Tables 5 and 6. In the AB stacking, the bottom and top layers exhibit net Bader charges of +0.0051 $e$ and -0.0051 $e$, respectively, signifying charge transfer from the top to the bottom layer. This confirms a negative charge center in the bottom layer and polarization along the +z direction. Conversely, the Bader charges of the bottom and top layers of BA stacking are -0.0051 $e$ and +0.0051 $e$, respectively, giving rise to a FE polarization oriented along the -z direction.

	We further confirm the magnetic ground states of both NP and FE phases by considering FM, AFM1 (A-type AFM---intralayer FM and interlayer AFM), and five other AFM orderings (AFM2-6, see Supplementary Fig. 4). Among them, the AFM1 is found to be energetically favored for both cases (see Supplementary Table 7). More importantly, while the NP phase exhibits zero net magnetization, the FE AB and BA stackings show a sizable net magnetic moment of up to 0.19~$\mu_\mathrm{B}$ per unit cell with opposite signs, as shown in Fig.~\ref{fig:fig2}c. All atomic magnetic moments for both the NP and FE phases are listed in Supplementary Table 8. More details about the exchange parameters and Curie temperature of the FE FiM Fe$_5$GeTe$_2$ bilayer, in comparison with the FM Fe$_5$GeTe$_2$ monolayer, are provided in Supplementary Section B.

\begin{figure}
		\includegraphics[width=1\columnwidth]{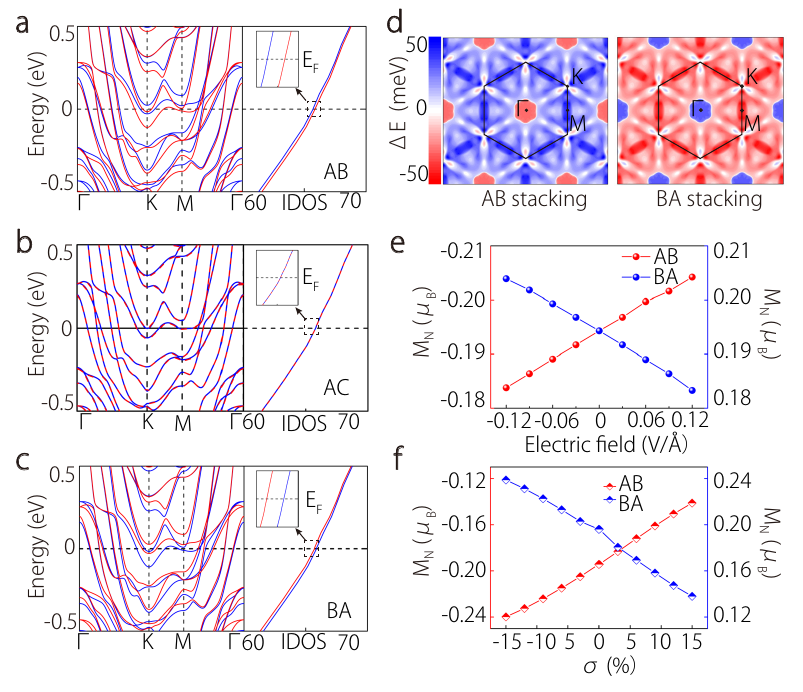}
		\caption{\textbf{Electronic structures and magnetoelectric responses of Fe$_5$GeTe$_2$ bilayer.} \textbf{a-c} Band structures and $iDOS$ of the AB, AC, and BA stackings. Red and blue denote spin-down and spin-up channels, respectively. \textbf{d} Spin-splitting maps (spin-up minus spin-down) of two selected bands crossing the Fermi energy for the AB and BA stackings. \textbf{e, f} Dependence of $M_N$ on  the applied out-of-plane electric field and the interlayer distance variation $\sigma$ in the AB- and BA-stacked Fe$_5$GeTe$_2$ bilayers, respectively.}
		\label{fig:fig3}
	\end{figure}

	To gain deeper insight into the origin of this sizable net magnetization, we then analyze the spin-resolved electronic structure and its evolution across the FE transition. As shown in Fig.~\ref{fig:fig3}b, the NP phase with the AC stacking exhibits nonrelativistic spin-degenerate band structures, with several bands crossing the Fermi level along the high-symmetry paths of the Brillouin zone (BZ). This spin degeneracy is protected by the [$\mathcal{C}_{2}{\parallel}\mathcal{M}_z$] symmetry that connects the two spin sublattices. Accordingly, the $iDOS$ for spin-up and spin-down channels remains equal ($iDOS^\uparrow(E_F) = iDOS^\downarrow(E_F)$), resulting in a vanishing net magnetic moment as described by Eq.~\eqref{eq:moment}. In contrast, for the FE phase with AB stacking, the interlayer sliding breaks the [$\mathcal{C}_{2}{\parallel}\mathcal{M}_z$] symmetry. As a result, the two spin sublattices are no longer connected by any symmetry operation, leading to spin-splitting electronic bands throughout the BZ. Given the metallic nature of the system, this spin splitting induces a polarization in the spin-resolved $iDOS$ below the Fermi level, with $iDOS^\uparrow(E_F) < iDOS^\downarrow(E_F)$ (see Fig. \ref{fig:fig3}a), which gives rise to a net magnetic moment of -0.19 $\mu_\mathrm{B}$ per unit cell. As detailed in Supplementary Table 8, the spin-up Fe1$-$Fe5 atoms possess smaller local magnetic moments compared to the spin-down Fe6$-$Fe10 atoms, confirming the FiM character of the FE state.
	
	For the other FE phase with BA stacking, it can be regarded as being derived from the AB stacking by applying a combined  operation [$\mathcal{C}_{2}{\parallel}\mathcal{M}_z$]. According to Eq.~\eqref{eq:PT}, this FE switching operation simultaneously reverses both $P_z$ and $S$ without altering the N{\'e}el vector. The projected 2D spin-splitting maps (see Fig. \ref{fig:fig3}d) clearly illustrate the opposite $S$ of the AB and BA stackings, consistent with the reversal of the spin-resolved band structures induced by FE switching (Fig. \ref{fig:fig3}c). As a consequence, the spin-resolved $iDOS$ below the Fermi level becomes inverted, with $iDOS^\uparrow(E_F) > iDOS^\downarrow(E_F)$ in contrast to the AB case. This reversal leads to an opposite net magnetic moment direction, while maintaining the same magnitude (see Fig. \ref{fig:fig2}c).

	Furthermore, we examine the effect of an external electric field on the magnetic behavior of the FE FiM phase.  As depicted in Fig. \ref{fig:fig3}e, the net magnetization exhibits an approximately linear response to a small external electric field. The corresponding magnetoelectric coupling coefficient can be evaluated using~\cite{duanSurfaceMagnetoelectricEffect2008,liuMagnetoelectricCouplingMultiferroic2020a}
	\begin{equation}
	\alpha \approx \mu_0 \frac{\Delta M}{E},
    \end{equation}
where $\mu_0$ denotes the vacuum permeability, $\Delta M$ is the change in magnetization, and $E$ is the applied electric field. Based on this expression, we estimate the linear magnetoelectric coupling coefficient to be approximately $7.13 \times 10^{-14}$G$\cdot$cm$^2$/V for both AB- and BA-stacked FE phases. This value markedly exceeds those reported in conventional systems such as Fe thin films ($2.9 \times 10^{-14}$G$\cdot$cm$^2$/V)~\cite{duanSurfaceMagnetoelectricEffect2008}. Remarkably, it is comparable to that of the prototypical sliding FE fc-FiM bilayer VS$_2$, which reaches a similar value ($9.8 \times 10^{-14}$G$\cdot$cm$^2$/V) when the interlayer distance is artificially reduced by 0.6 \AA ~\cite{liuMagnetoelectricCouplingMultiferroic2020a}. These results underscore the strong magnetoelectric response inherent in this FE FiM metallic system, even without external structural tuning.

		\begin{figure}
		\includegraphics[width=1.0\columnwidth]{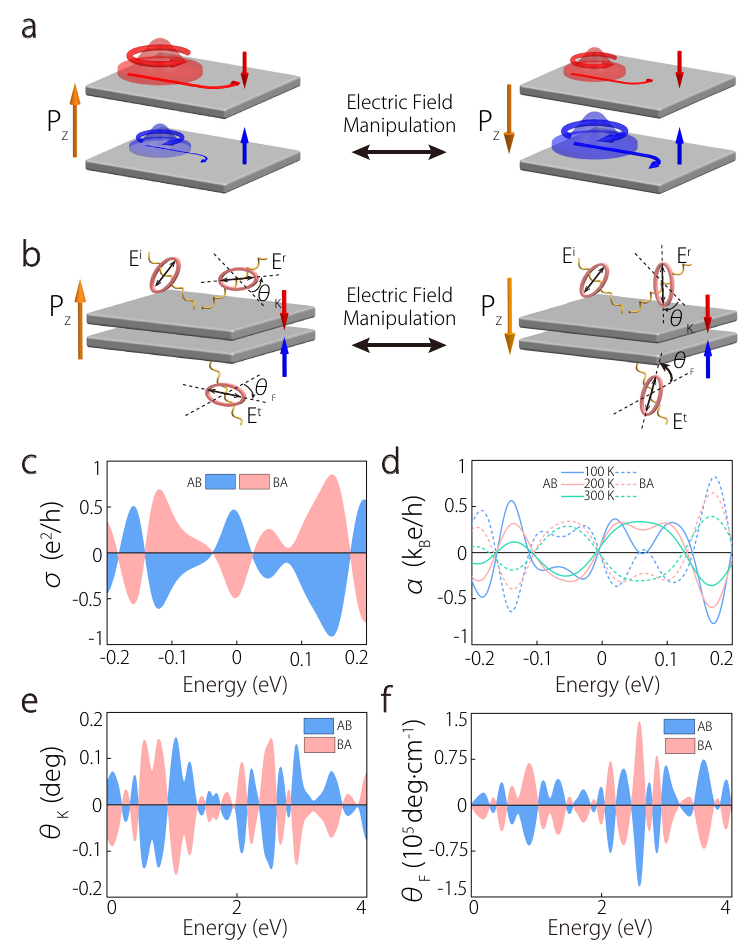}
		\caption{\textbf{Sign-reversible anomalous transport properties for Fe$_5$GeTe$_2$ bilayer.} \textbf{a} Schematic illustration of sign reversal in the AHE and ANE across the FE transition. \textbf{b} Schematic illustration of sign reversal in the MO Kerr and Faraday rotation angles across the FE transition. \textbf{c} AHCs of the AB and BA stackings. \textbf{d} ANCs of the AB and BA stackings at different temperatures. Solid and dashed lines correspond to AB and BA stackings, respectively. \textbf{e, f} MO Kerr and Faraday rotation angles for the AB and BA stackings.}
		\label{fig:fig4}
	\end{figure}

	Moreover, the interlayer coupling also plays a significant role in modulating the net magnetization of the system. To quantify this effect, the interlayer distance was systematically varied by $\pm 15\%$ with respect to the equilibrium value ($d_{0} = 3.09~\text{\AA}$) to simulate interlayer compression and expansion, i.e., $d = d_{0}(1 \pm \sigma)$, where $\sigma \in [-15\%, +15\%]$ (see Supplementary Fig.~6). Within this range, we first examine the magnetic ground state of the AB-stacked FE phase under different interlayer distances. As shown in Supplementary Fig. ~7a, the energy difference between the FM and AFM1 configurations ($\Delta E = E_{\mathrm{FM}} - E_{\mathrm{AFM1}}$, in meV per unit cell) remains positive, confirming that the AFM1 configuration is consistently the magnetic ground state across all interlayer distances. Furthermore, as illustrated in Fig.~\ref{fig:fig3}f, upon varying the interlayer distance, the two FE phases exhibit opposite magnetic moments with nearly identical magnitudes. In both phases, reducing the interlayer distance strengthens the interlayer coupling, which in turn enhances the net magnetic moment. This behavior demonstrates that the magnetoelectric coupling in Fe$_5$GeTe$_2$ bilayer can be effectively tuned by the interlayer interaction strength.

Notably, sliding ferroelectricity has been experimentally realized, first observed in few-layer WTe$_2$ in 2018~\cite{feiFerroelectricSwitchingTwodimensional2018b}, where switchable interlayer polarization was confirmed by conductance hysteresis and piezoresponse measurements. The demonstrated electric-field-driven reversal of polarization establishes a clear precedent for realizing sliding ferroelectricity in van der Waals metals. Given the similarly layered structure of Fe$_5$GeTe$_2$, our proposed sliding FE FiM metal could be experimentally accessible with current techniques \cite{sharmaRoomtemperatureFerroelectricSemimetal2019,xiaoBerryCurvatureMemory2020,yasudaStackingengineeredFerroelectricityBilayer2021,wuSlidingFerroelectricity2D2021,westonInterfacialFerroelectricityMarginally2022a}, making its experimental validation highly probable in the near future.

\vspace{0.3cm}	
\noindent{\bf Considerable sign-reversible anomalous transport properties}

\noindent
 Owing to its metallic nature, the system may exhibit spontaneous transport responses, particularly the AHE and ANE, which are typically negligible or entirely absent in conventional AFM insulators within the band gap. These anomalous transport responses are highly dependent on the N{\'e}el vector orientation or specific magnetic symmetries~\cite{XD-Zhou2021,XD-Zhou2024,JQ-Feng2024}. As summarized in Supplementary Section C, both FE and NP phases exhibit in-plane magnetic anisotropy with the small magnetic anisotropy energy ($<$ 1 meV), which implies that the magnetization orientation is highly sensitive to experimental conditions and can be easily reoriented by an external magnetic field or through spin-orbit torque effects~\cite{dengGatetunableRoomtemperatureFerromagnetism2018a,tanGateControlledMagneticPhase2021}. For example, a magnetization reorientation transition between in-plane and out-of-plane magnetization has been experimentally demonstrated in four-layer Fe$_3$GeTe$_2$ \cite{dengGatetunableRoomtemperatureFerromagnetism2018a}, further confirming the ease of magnetic axis manipulation arising from the small magnetic anisotropy energy revealed in our calculations.
	
We conducted detailed magnetic symmetry analyses to determine symmetry-imposed anomalous transport responses in both the NP and FE phases, as summarized in Supplementary Tables~13 and 14 as well as Section D. For the transient NP phase, anomalous transport responses may arise only when the N{\'e}el vector possesses an in-plane component, whereas they vanish for an out-of-plane orientation---a characteristic feature of magnets protected by the combined $[\mathcal{C}_2||\mathcal{M}_z]$ symmetry~\cite{baiAnomalousHallEffect2025a} (see Supplementary Fig. 9, Table S13, and  Section D for more details). For the ground-state FE phase, anomalous transport responses are forbidden when the N{\'e}el  vector lies along the in-plane $\langle100\rangle$ or $\langle110\rangle$ crystallographic direction families, while they are allowed for all other orientations (see Supplementary Table 14). Specifically, when the N{\'e}el vector rotates within the $x$-$y$ plane, anomalous transport responses periodically appear and disappear as the magnetic point group alternates between the symmetry-allowed ($m^\prime$, $1$) and symmetry-forbidden ($m$) configurations. A similar symmetry evolution occurs in the $z$-$x$ plane, where the anomalous transport responses are depressed when the N{\'e}el vector is aligned with the $\pm x$ axis ($\langle100\rangle$ direction), but are allowed for other directions, e.g., the out-of-plane direction associated with the magnetic point group $3m'$. As representative examples, the main text focuses on the anomalous transport properties of the FE phase with the magnetization oriented out-of-plane. The cases with in-plane N{\'e}el-vector orientations are presented in Supplementary Figs. 10 and 11.

	As shown in Figs. \ref{fig:fig4}c-f, for the AB-stacked FE phase, the anomalous Hall conductivity (AHC) reaches approximately 0.49 $e^2/h$ at the Fermi level, approaching the half-quantized value. The anomalous Nernst conductivity (ANC) near the Fermi level attains a peak of 0.36 $k_B e/h$ ($\sim$0.62 AK$^{-1}$·m$^{-1}$) at 100 K and 0.34 $k_B e/h$ ($\sim$0.57 AK$^{-1}$·m$^{-1}$) at 300 K, exceeding that of noncollinear antiferromagnet Mn$_3Y$ ($Y$ = Ge, Ga, Sn) ($\sim$0.3 AK$^{-1}$·m$^{-1}$)~\cite{ikhlasLargeAnomalousNernst2017}. Moreover, the Kerr rotation angle reaches 0.14 deg, which is significantly larger than that of Mn$_3Y$ ($\sim$0.02 deg)~\cite{higoLargeMagnetoopticalKerr2018}, while the Faraday rotation reaches 1.36 $\times$ 10$^5$ deg$\cdot$cm$^{-1}$, surpassing that of FM Y$_3$Fe$_5$O$_{12}$ ($\sim$0.02 $\times$ 10$^5$ deg$\cdot$cm$^{-1}$)~\cite{boudiarMagnetoopticalPropertiesYttrium2004,tomitaMagnetoOpticalKerrEffects2006}. As summarized in Supplementary Section C, the calculated anomalous transport properties are well converged. Furthermore, owing to the reversal of spin splitting $S$ across the FE transition, the anomalous transport properties can be controlled by an external electric field (see Fig. \ref{fig:fig4}a and b). Specifically, the AB- and BA-stacked FE phases are connected by $\mathcal{TM}_{z}$ symmetry under which the anomalous transport properties are odd, leading to identical magnitudes but opposite signs in the AHC, ANC, as well as the Kerr and Faraday rotation angles  (see Fig. \ref{fig:fig4}c-f). This property allows for the nonvolatile control of these anomalous transport responses through electric field-driven FE switching, which is highly desirable for future spintronic applications. Finally, we emphasize that the anomalous transport properties are protected by symmetry. Variations in the interlayer distance only slightly affect their magnitudes but do not alter their qualitative existence, as detailed in Supplementary Section E.

 To summarize, based on AFM metallic bilayers with spin degeneracy protected by $[\mathcal{C}_2||\mathcal{M}_z]$ symmetry, we propose a design strategy for 2D sliding FE FiM metals that achieve triply-coupled switching, in which FE transitions simultaneously reverse FE polarization, net magnetization, and spin splitting. This enables strong magnetoelectric coupling and electric control of pronounced anomalous transport properties. Taking bilayer Fe$_5$GeTe$_2$ as an example, the AC stacking behaves as an AFM NP metal. During FE switching, interlayer sliding breaks the [$\mathcal{C}_2{\parallel}\mathcal{M}_z$] symmetry, decouples the two spin sublattices, and induces spin splitting. Owing to its metallic band structure, the FE transition leads to unequal spin-resolved integrated densities of states at the Fermi level, triggering a transition into two FE FiM phases characterized by reversed FE polarization, net magnetization, and spin splitting. The resulting spin-polarized metallic states give rise to pronounced anomalous transport properties, including the AHE, ANE, and MOEs. Furthermore, the FE phase exhibits sign-reversible anomalous transport near the Fermi level that is controllable by an electric field through FE switching. These findings open a new avenue for designing 2D multiferroic materials that integrate FE, FiM, and metallic properties for next-generation nanoelectronic and spintronic applications.

	\vspace{0.3cm}
	\noindent{\Large \bf Methods}
	
	\vspace{0.05cm}
	\noindent{\bf First-principles calculations} 
	
	\noindent
	The first-principles calculations were performed using the Vienna ab initio simulation package (VASP)~\cite{kresseInitioMoleculardynamicsSimulation1994a,kresseEfficientIterativeSchemes1996b} with generalized gradient approximation (GGA) of the Perdew-Burke-Ernzerhof (PBE) functional~\cite{perdewGeneralizedGradientApproximation1996a}. The projector-augmented wave (PAW) method~\cite{blochlProjectorAugmentedwaveMethod1994a} was adopted and the cutoff energy was set as 500 eV. The first Brillouin zone (BZ) was sampled with a Monkhorst-Pack $k$-mesh of $9\times 9\times 1$. As suggested by previous studies~\cite{tanGateControlledMagneticPhase2021,ershadradUnusualMagneticFeatures2022,xieStrainInducedInterlayerMagnetic2023a,ghoshUnravelingEffectsElectron2023}, the GGA functional without applying a Hubbard U term provides a more reliable description of ground-state electronic and magnetic properties of the Fe$_5$GeTe$_2$ system. For the calculation of the out-of-plane FE polarization and magnetocrystalline anisotropy energy (MAE), a denser $15\times15\times1$ Monkhorst-Pack $k$-mesh was employed. The energy and force convergence criteria were set as 10$^{-6}$ eV and 0.01 eV/\AA, respectively. A vacuum layer with a thickness of about 18 \AA  \ was taken to avoid artificial interactions between adjacent slabs and the interlayer van der Waals interaction was treated based on the DFT-D3 method of Grimme with zero-damping function~\cite{grimmeConsistentAccurateInitio2010,grimmeEffectDampingFunction2011}. The dipole correction was applied along the $z$ direction to eliminate the spurious interaction between periodic images and the FE transition pathway was obtained by the climbing image nudged elastic band (CL-NEB) method~\cite{henkelmanClimbingImageNudged2000,smidstrupImprovedInitialGuess2014}. To investigate the anomalous transport properties, the maximally localized Wannier functions (MLWFs) for the $s$-, $p$-, $d$-orbitals of Fe, Ge and Te atoms in Fe$_5$GeTe$_2$ were constructed by using the Wannier90~\cite{souzaMaximallyLocalizedWannier2001a,mostofiWannier90ToolObtaining2008}, the intrinsic anomalous Hall conductivity (AHC) and anomalous Nernst conductivity (ANC) were calculated using the WannierTools package~\cite{wuWannierToolsOpensourceSoftware2018a}.

	\vspace{0.3cm}
	\noindent{\bf Anomalous hall conductivity and anomalous nernst conductivity} 

	\noindent
The intrinsic AHC and the ANC were calculated on an ultra-dense $k$-mesh $251 \times 251 \times 1$ using the Berry phase theory~\cite{yaoFirstPrinciplesCalculation2004b},
\begin{equation}
\sigma_{xy} = -\frac{e^2}{\hbar} \sum_n \int \frac{d^2k}{(2\pi)^2} \, \Omega^n_{xy}(\mathbf{k}) f_{n\mathbf{k}},
\end{equation}

\begin{equation}
\begin{aligned}
\alpha_{xy}^{A}
= \frac{e}{\hbar T} \sum_{n} \int \frac{d^{2}k}{(2\pi)^{2}} \,
\Omega^{n}_{xy}(\mathbf{k}) \times 
\Big[
(\varepsilon_{n\mathbf{k}} - \mu)\, f_{n\mathbf{k}} \\
\quad +\, k_{B}T \ln\!\left(1 + e^{-(\varepsilon_{n\mathbf{k}}-\mu)/k_{B}T}\right)
\Big].
\end{aligned}
\label{eq:ANC}
\end{equation}
where $n$, $\mathbf{k}$, $T$, $\mu$, and $k_B$ are band index, crystal momentum, temperature, chemical potential, and Boltzmann constant, respectively. ${(x, y)}$ denote the Cartesian coordinates. $f_{n\mathbf{k}} = 1/[\exp((\varepsilon_{n\mathbf{k}} - \mu)/k_BT) + 1]$ is the Fermi-Dirac distribution. $\Omega^n_{xy}(\mathbf{k})$ is the band-resolved Berry curvature, given by:
\begin{equation}
\Omega^n_{xy}(\mathbf{k}) = -\sum_{n' \neq n} \frac{2\, \text{Im} \left[ \langle \psi_{n\mathbf{k}} | \hat{v}_x | \psi_{n'\mathbf{k}} \rangle \langle \psi_{n'\mathbf{k}} | \hat{v}_y | \psi_{n\mathbf{k}} \rangle \right]}{(\omega_{n'\mathbf{k}} - \omega_{n\mathbf{k}})^2}.
\end{equation}
Here, $\hat{v}_x, \hat{v}_y$ are the velocity operators, and $\psi_{n\mathbf{k}}$ $(\hbar\omega_{n\mathbf{k}}= \varepsilon_{n\mathbf{k}})$ is the eigenvector (eigenvalue) at band index $n$ and momentum $\mathbf{k}$.

\vspace{0.3cm}
\noindent{\bf Optical conductivity} 

\noindent
The optical conductivity is calculated via the Kubo-Greenwood formula~\cite{mostofiWannier90ToolObtaining2008,yatesSpectralFermiSurface2007}:
\begin{equation}
\sigma_{\alpha\beta} = \frac{ie^2 \hbar}{N_k \Omega_c} \sum_{\mathbf{k}} \sum_{n,m} \frac{f_{m\mathbf{k}} - f_{n\mathbf{k}}}{\varepsilon_{m\mathbf{k}} - \varepsilon_{n\mathbf{k}}} 
\frac{\langle \psi_{n\mathbf{k}} | \hat{v}_\alpha | \psi_{m\mathbf{k}} \rangle \langle \psi_{m\mathbf{k}} | \hat{v}_\beta | \psi_{n\mathbf{k}} \rangle}
{\varepsilon_{m\mathbf{k}} - \varepsilon_{n\mathbf{k}} - (\hbar\omega + i\eta)},
\end{equation}
where $\alpha, \beta$ denote Cartesian directions ${(x, y)}$, $N_k$ is the total number of $k$-points for sampling the BZ (here, an ultra-dense $k$-mesh $251 \times 251 \times 1$ was used), $\Omega_c$ is the unit cell volume, $\hbar\omega$ is the photon energy, and $\eta$ is the broadening parameter.

For topologically trivial systems, the complex magneto-optical Kerr and Faraday angles can be described as~\cite{suzukiNewMagnetoopticalTransition1992,guoBandtheoreticalInvestigationMagnetooptical1995a,ravindranMagneticOpticalMagnetooptical1999,fengTunableMagnetoopticalEffects2016},
\begin{equation}
\phi _K = \theta_K + i\varepsilon_K = i\frac{2\omega d}{c} \frac{\sigma_{xy}}{\sigma^s_{xx}},
\end{equation}
\begin{equation}
\phi _F = \theta_F + i\varepsilon_F = \frac{\omega d}{2c} (n_+ - n_-),
\end{equation}

Here, $\theta_{K,F}$ and $\varepsilon_{K,F}$ are the Kerr (Faraday) rotation angle and ellipticity. 
$c$, $\omega$, and $d$ are the speed of light in vacuum, the frequency of incident light, and the effective thickness of the thin film, respectively. $\sigma_{xx}$ and $\sigma_{xy}$ are the diagonal and off-diagonal components of the optical conductivity tensor for the magnetic thin film. 
$\sigma_{xx}^{s} = i(1 - n^2)\omega / 4\pi$ represents the diagonal optical conductivity of a nonmagnetic substrate (commonly SiO$_2$). 
The eigenvalues of the dielectric tensor of the magnetic film are given by 
$n_{\pm}^2 = 1 + \frac{4\pi i}{\omega} (\sigma_{xx} \pm i\sigma_{xy})$.

\vspace{0.3cm}
\noindent{\Large \bf Data availability}

\noindent
The data that support the findings of this study are available from the corresponding author upon reasonable request.

\vspace{0.3cm}
\def\bibsection{\Large \noindent{\bf References}}	
\bibliographystyle{naturemag.bst}	
\bibliography{ref}

\vspace{0.3cm}
\noindent{\Large \bf Acknowledgements}

\noindent
The authors thank Dr. Xiuxian Yang and Prof. Zhi-Ming Yu for valuable and insightful discussions. This work was supported by the National Key R\&D Program of China (Grant No. 2022YFA1402600), the Australian Research Council Discovery Early Career Researcher Award (Grant No. DE240100627), the Australian Research Council Discovery Project (Grant No. DP260102992), the National Natural Science Foundation of China (Grants No. 12304066), and the Basic Research Program of Jiangsu (Grants No. BK20230684). We acknowledge the allocation of high-performance computer time from UOW Partner Share. We also acknowledge the computational resources from the National Computational Infrastructure (NCI), which were allocated from the National Computational Merit Allocation Scheme supported by the Australian Government.

\vspace{0.3cm}
\noindent{\Large \bf Author contributions}

\noindent
X.W. led the project, with X.Z. and Z.C. contributing to the conceptualization. Z.G., S.Q., and X.Z. performed the symmetry analysis and first-principles calculations on ferroelectric ferrimagnetic metals. Z.G. and X.Z. carried out the symmetry analysis and calculations of anomalous transport properties. Z.G., X.Z., and X.W. prepared the manuscript with input from S.Q., Z.C., and W.W..  Z.G. and S.Q. contributed equally to this work.

	\vspace{0.3cm}
	\noindent{\Large  \bf Competing interests}
	
	\noindent{The authors declare no competing interests.}
	
	\vspace{0.3cm}
	\noindent{\Large  \bf Additional information}
	
	\noindent{{\bf Supplementary information} is available for this paper at https://doi.org/xxxxxx.}
	
	\vspace{0.3cm}
	\noindent{{\bf Reprints and permission information} is available at http://www.nature.com/reprints}
	
	\vspace{0.3cm}
	\noindent{{\bf Publisher's note} Springer Nature remains neutral with regard to jurisdictional claims in published maps and institutional affiliations.}

\end{document}